\documentclass[a4paper]{jpconf}
\usepackage{graphicx}
\usepackage{xspace}
\usepackage{mydefs}
\begin{document}
\title{BSM constraints from model-independent measurements: A {\sc Contur} Update}

\author{Jon Butterworth}

\address{Department of Physics \& Astronomy, UCL, London, Gower St., London WC1E 6BT, UK}

\ead{j.butterworth@ucl.ac.uk}

\begin{abstract}
Particle-level measurements, especially of differential cross-sections, made in fiducial regions of phase-space have a 
high degree of model-independence and can therefore be used to give information about a wide variety of 
Beyond the Standard Model (BSM) physics implemented in Monte Carlo generators, using a broad range of final states.
The \contur package is used to make such comparisons.
We summarise a snapshot of current results for a number of BSM scenarios; 
a UV complete model in which the global Baryon-number minus Lepton-number symmetry is gauged;
several Dark Matter simplified models, and two generic light scalar models.
\end{abstract}

\section{Introduction}

The \contur programme was presented in Ref.~\cite{Butterworth:2016sqg}. It makes use of the 
well-developed chain of phenomenological and experimental software tools which allows a new model to
be coded in FeynRules\cite{Alloul:2013bka}, exported in a standard format~\cite{Degrande:2011ua} 
intelligible to full final-state 
event generators~\cite{Buckley:2011ms} such as Herwig~\cite{Bellm:2015jjp}, 
and compared to particle-level collider data stored in 
HEPDATA~\cite{Maguire:2017ypu} using the analyses encoded in Rivet~\cite{Buckley:2010ar}. 
Essentially the idea is to look at the predicted contributions from a BSM model to the fiducial phase space
of measurements which have been shown to be in good agreement with the Standard Model (SM), and see if
the additional contributions would have been visible in the measurement. If so, that model (or those 
parameters of that model) are said to be disfavoured by the data, at some confidence level. More details 
are given in~\cite{Butterworth:2016sqg}.
In each case, the Herwig event generator~\cite{Bellm:2015jjp,Bellm:2017bvx} is used to generate inclusively all signatures 
involving the new particle content of the model.
In this presentation we give an update of some recent results.

\section{Gauged B-L model with heavy neutrinos}

There is significant interest in extensions to the SM in which the global symmetry behind the conservation of $B-L$ 
(Baryon number minus lepton number) is gauged, giving an additional $U(1)_{B-L}$ symmetry and an associated new gauge boson. 
One such model was discussed in \cite{Deppisch:2018eth}, in which the additional $U(1)_{B-L}$ gauge symmetry is 
spontaneously broken by an extra singlet Higgs. To make the model anomaly-free it also incorporates three generations of 
neutral leptons sterile under the SM gauge interactions, thereby enabling the Seesaw mechanism of light neutrino mass generation. 

\begin{figure}[h]
\includegraphics[width=0.9\textwidth]{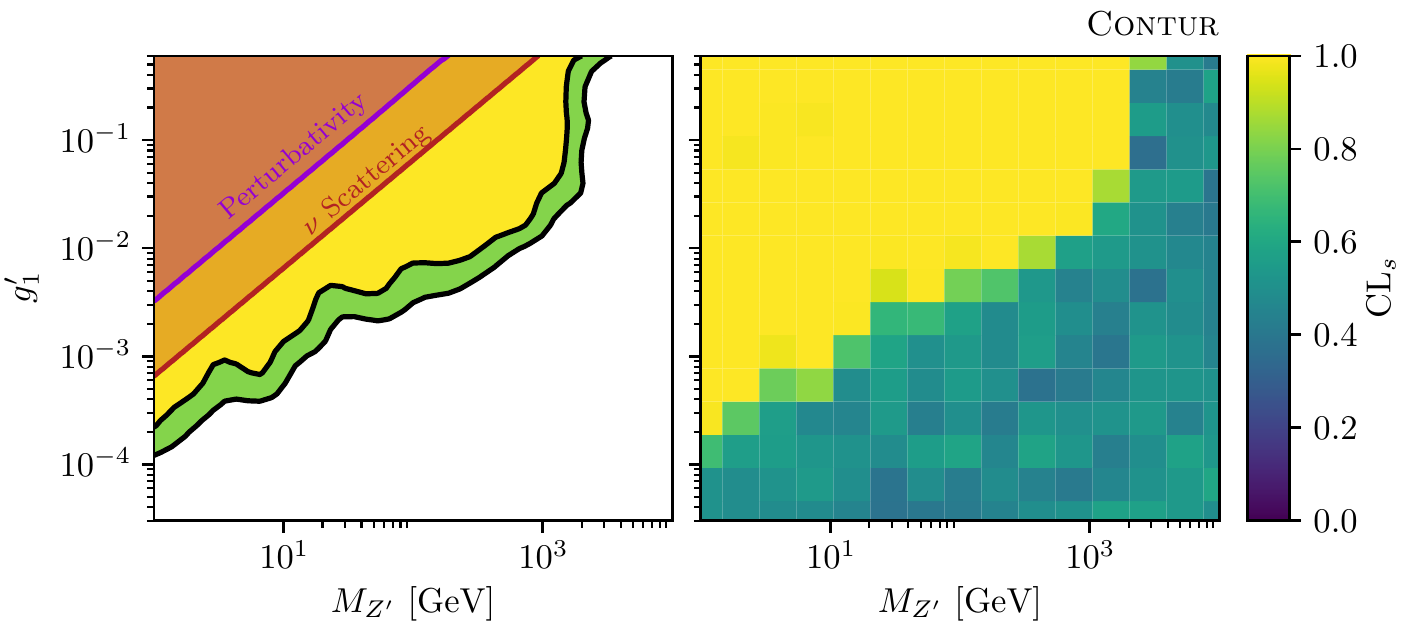}
\caption{\label{fig:caseC}Sensitivity of LHC measurements to the BSM contribution from a gauged B-L model in the \MZP vs \gp plane.
Left, 95\% (yellow) and 68\% (green) excluded contours. Right, underlying heatmap of exclusion at each scanned parameter space point. 
$\SA = 0.2, \MHH = 200$~GeV; theory bounds and constraints from $M_{W}$ and from neutrino scattering cross sections are also shown.}
\end{figure}

In \cite{Amrith:2018yfb}, the potential signatures from a range of model parameters and processes were considered. While the heavy 
neutrinos may give rare and distinct signals (as discussed in \cite{Deppisch:2018eth}), they play no significant role in
the signatures addessed by \contur. The important parameters are the mass of the new gauge boson \MZP 
and its coupling, \gp, to the SM, and the mass of the new Higgs, \MHH, and its mixing, \SA, with the SM Higgs. If the Higgs
sector is decoupled ($\SA=0$), the model reduces a rather standard \ZPRM model, and the majority of the parameter space reachable 
by \contur is already excluded by dedicated searches. More interesting is the case where \SA is non-zero. An example, 
for $\MHH = 200$~GeV and $\SA = 0.2$, is shown in Figure~\ref{fig:caseC}. The \contur analysis of ATLAS, CMS and LHCb data disfavours
a substantial region of the plane.

\begin{figure}[h]
\includegraphics[width=0.9\textwidth]{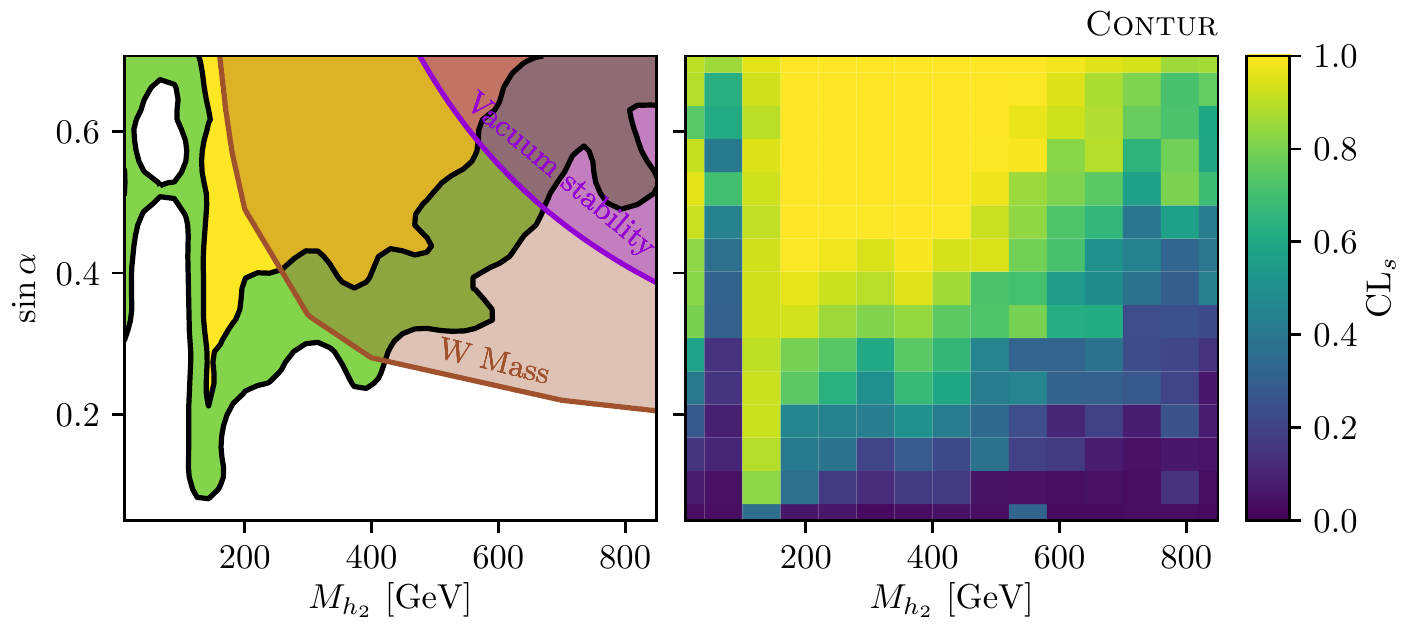}
\caption{\label{fig:caseE}Sensitivity of LHC measurements to the BSM contribution from a gauged B-L model in the \MHH vs \SA plane.
Legend as Figure~\ref{fig:caseC}, $\MZP = 35$~GeV, $\SA = 0.2$. The constraint from measurements of the $W$ mass is also shown.
}
\end{figure}

In regions where signatures involving direct production of the \ZPRM are not visible, either because it is too 
massive or \gp is too low, it is useful to
scan the \MHH vs \SA plane to see where signatures involving the BSM Higgs might play a role. An example is given in 
Figure~\ref{fig:caseE}. The constraints from $W$ mass measurements (where the additional Higgs can contribute to loop corrections)
are stringent. The \contur analysis extends them slightly, 
and there is lower sensitivity beyond this, apparent in the heatmap, showing that future measurements should be able to 
probe lower \SA values.

\section{Dark Matter}

\begin{figure}[h]
\includegraphics[width=0.9\textwidth]{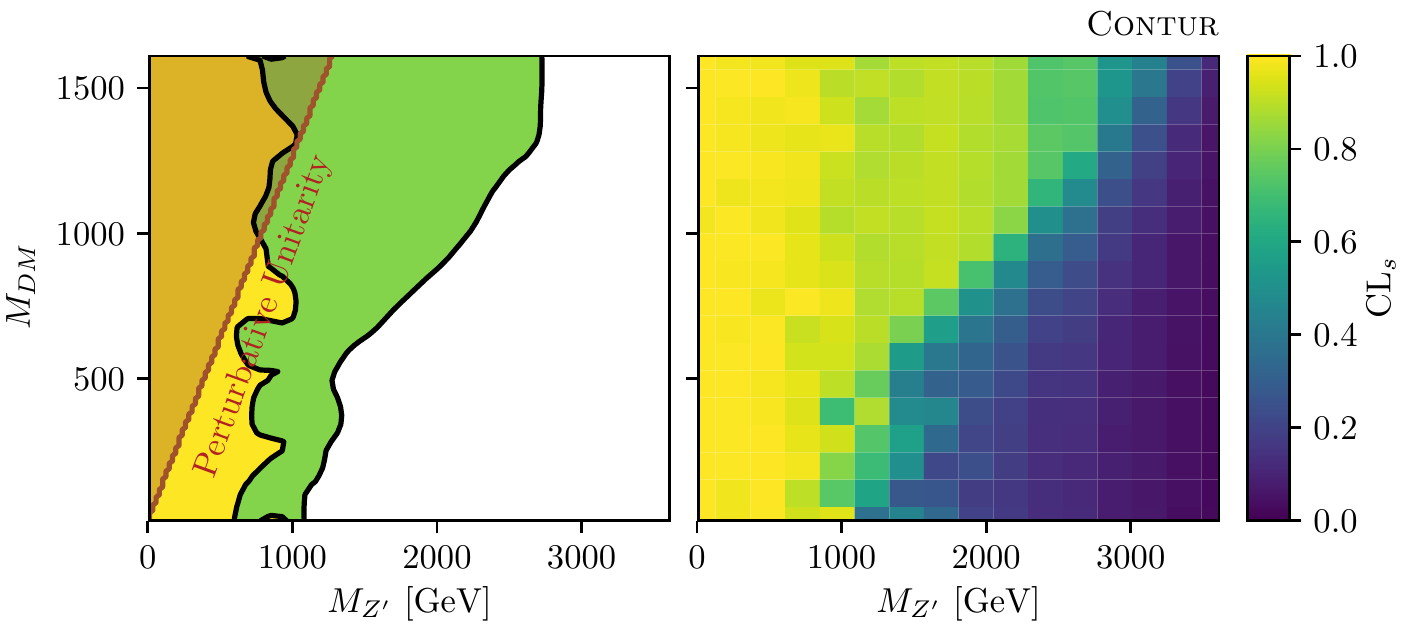}
\caption{\label{fig:dmlq} Sensitivity of LHC measurements to a Majorana DM candidate coupling to a spin-1 mediator \ZPRM, which
in turn couples to first generation quarks with coupling strength $\GQ = 0.25$.}
\end{figure}

A Dark Matter (DM) `simplified model' was considered in the first \contur paper~\cite{Butterworth:2016sqg}. In this model,
DM is a Majorana fermion which couples to a mediating spin-1 boson via an axial-vector current with strength \GDM. 
The boson in turn couples to first generation SM quarks with a vector coupling of strength \GQ. An update of the results for
$\GDM = 1, \GQ = 0.25$ is shown in Figure~\ref{fig:dmlq}. The data which have been added to HEPDATA and Rivet since 2016 lead to 
improved sensitivity, although the lack of the published 8~TeV and 13~TeV dijet data still weakens the analysis compared to searches 
with those datasets. It is hoped that these data will soon be available in Rivet. The diagonal structure along the line 
$\MZP \approx 2\MDM$ is caused by the reduced sensitivity when the \ZPRM decays to DM rather than to jets.

\begin{figure}[h]
\includegraphics[width=0.9\textwidth]{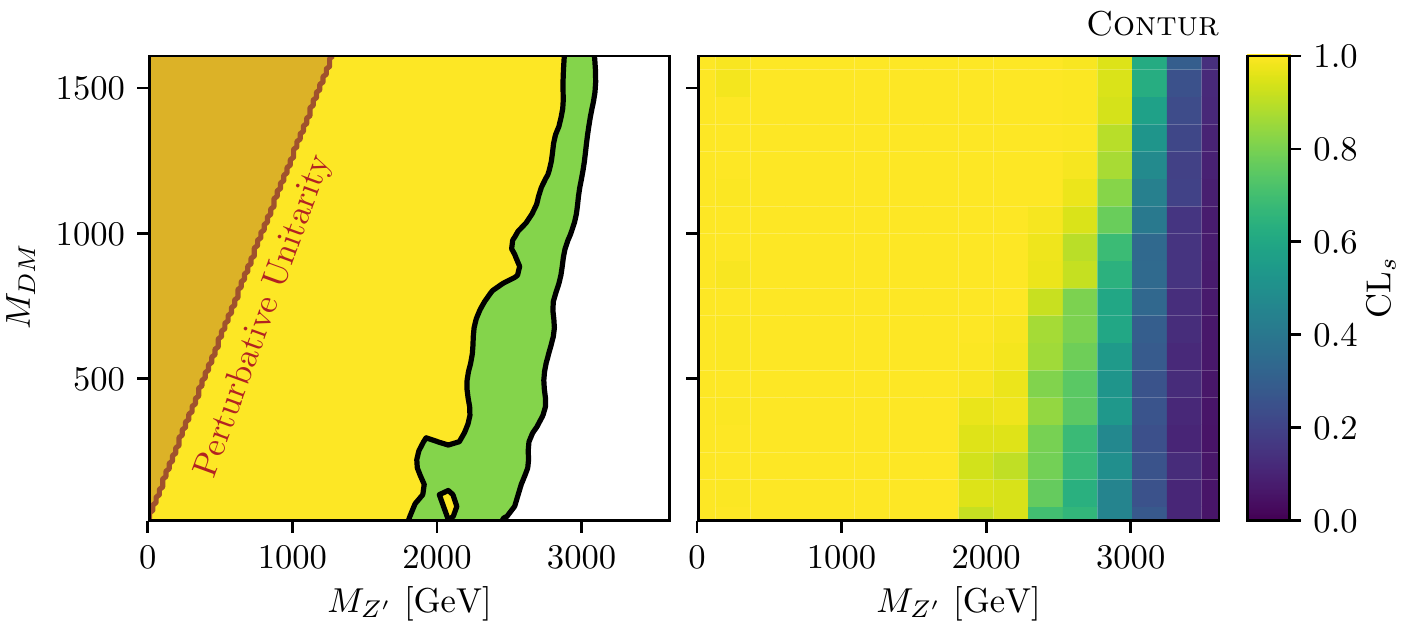}
\caption{\label{fig:dmhq} 
Sensitivity of LHC measurements to a Majorana DM candidate coupling to a spin-1 mediator \ZPRM, which
in turn couples to all three generations of quarks with coupling strength $\GQ = 0.25$.}
\end{figure}

\begin{figure}[h]
\includegraphics[width=0.9\textwidth]{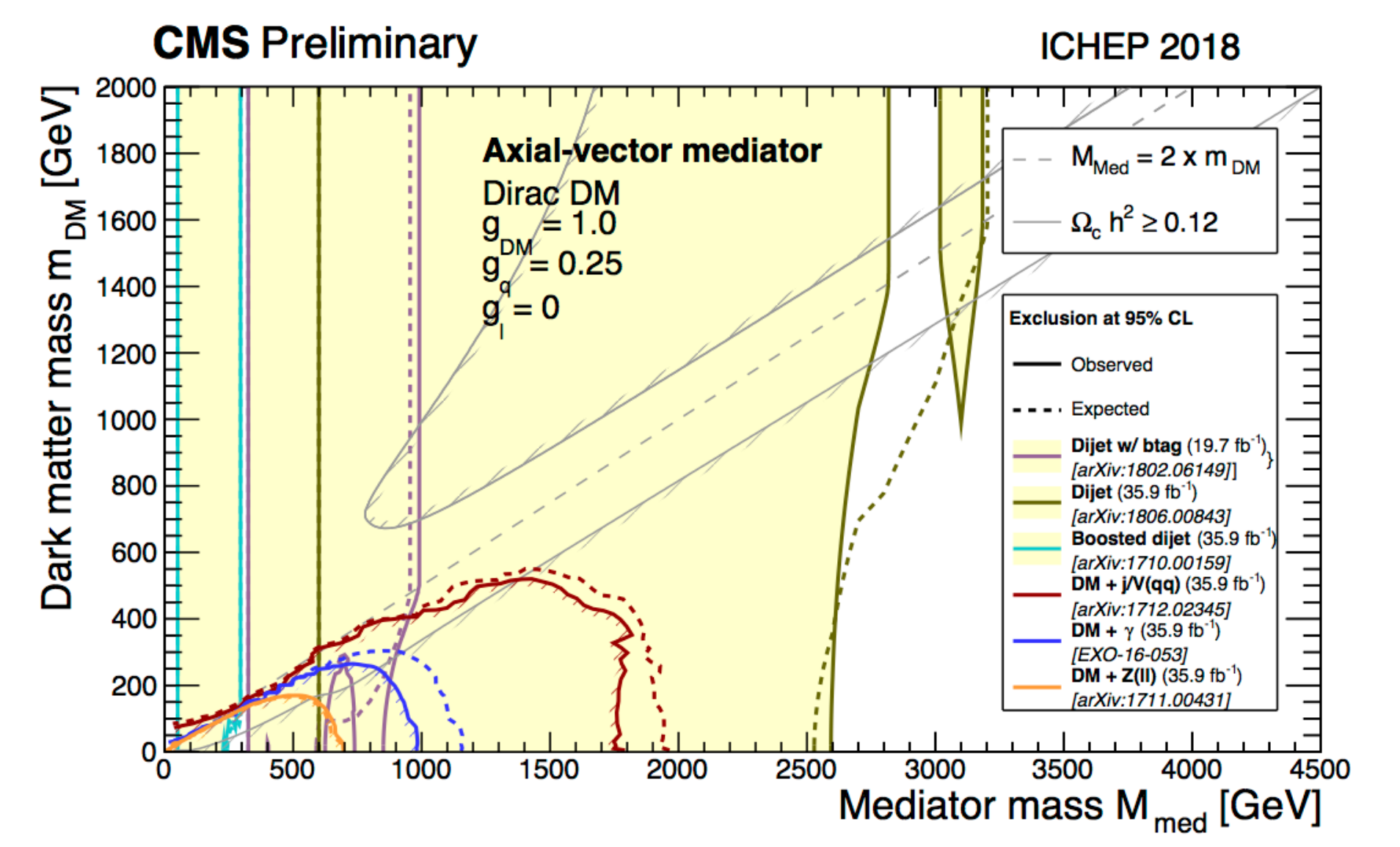}
\caption{\label{fig:CMS} 95\% CL observed and expected exclusion regions for dijet searches and missing energy based DM 
searches from CMS in the lepto-phobic Axial-vector model. 
Following the recommendation of the LHC DM working group \cite{Boveia:2016mrp,Albert:2017onk}
the exclusions are computed for a universal quark coupling of $\GQ = 0.25$ and $\GDM = 1.0$. 
It should also be noted that the absolute exclusion of the different searches as well as their relative importance, 
will strongly depend on the chosen coupling and model scenario. Therefore, the exclusion regions, relic density contours, 
and unitarity curve shown in this plot are not applicable to other choices of coupling values or model.}
\end{figure}

A simple modification of this model such that the \ZPRM couples to all three flavours of quarks has also been studied. The main 
impact of this change is to open up decays to top quark pairs once $\MZP > 2 M_{\rm top}$. Many top measurements from both ATLAS 
and CMS are available in Rivet for 8~TeV and 13~TeV data, so the lack of dijet data is less important. The improved sensitivity is 
clearly visible in Figure~\ref{fig:dmhq}.

\begin{figure}[h]
\includegraphics[width=0.9\textwidth]{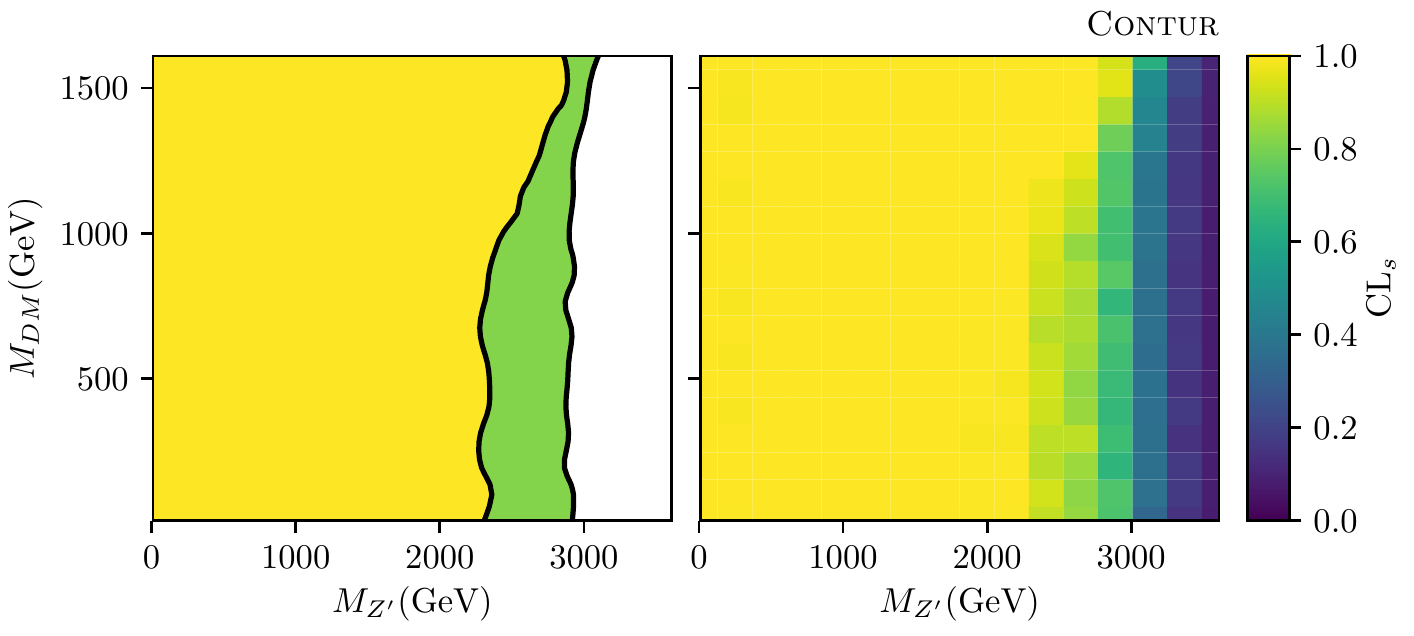}
\caption{\label{fig:dmsimpq} Sensitivity of LHC measurements to a Dirac DM candidate coupling to a spin-1 mediator \ZPRM, which
in turn couples to all three generations of quarks with coupling strength $\GQ = 0.25$.}
\end{figure}

This variant, with the \ZPRM coupling to all three generations, is much closer to the benchmark models studied by the experiments, 
for example in the summaries shown by CMS at ICHEP 2018 (see Figure~\ref{fig:CMS}), and the recent 
ATLAS compilation~\cite{ATLAS-CONF-2018-051}. The most
similar of these models has the same \ZPRM couplings, the only difference being that the DM candidate is now a Dirac fermion.

\begin{figure}[h]
\includegraphics[width=0.9\textwidth]{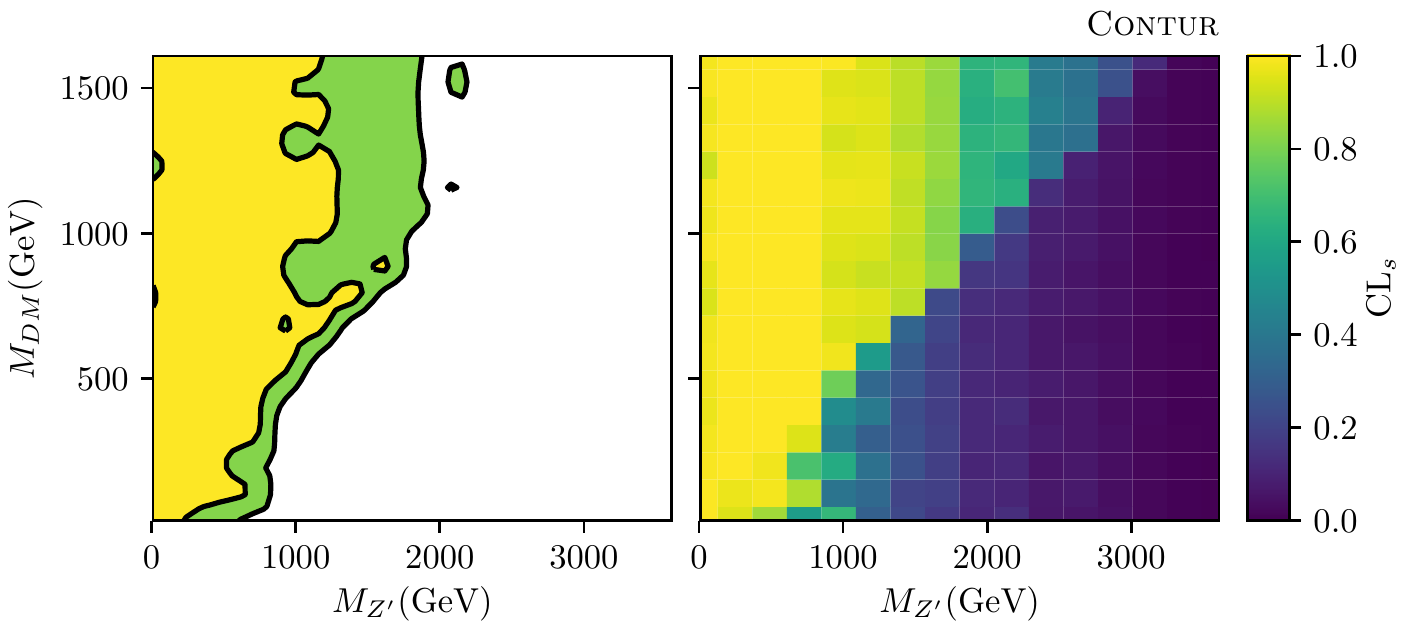}
\caption{\label{fig:dmsimpl} 
Sensitivity of LHC measurements to a Dirac DM candidate coupling to a spin-1 mediator \ZPRM, which
in turn couples to all three generations of quarks with coupling strength $\GQ = 0.1$ and to leptons with a 
coupling strength $\GL = 0.01$.}
\end{figure}

Exactly this model was also tested with \contur, with the result shown in Figure~\ref{fig:dmsimpq}. The \contur limits are
close to those given by the searches, despite the lack of dijet data and generally lower luminosity used in the measurements so far.

Another variant of the model, studied in \cite{ATLAS-CONF-2018-051}, reduces \GQ to 0.1 but allows a small coupling to
leptons, $\GL = 0.01$, opening up dilepton signatures. The \contur result for this scenario is shown in Figure~\ref{fig:dmsimpl}. 
The sensitivity is not as strong as that of the searches in  \cite{ATLAS-CONF-2018-051}, because in addition to the lack of 
8 and 13~TeV dijets, there is no 13~TeV dilepton measurement yet available in Rivet.

\section{Light Scalars}

Finally we consider LHC sensitivity to additional light scalar particles. These are a common feature in extensions of the SM, 
for example appearing in composite Higgs scenarios, or as the radion in models with extra dimensions~\cite{Angelescu:2017jyj}. 
Consideration of precision electroweak measurements, collider searches and flavour physics does not completely exclude the 
existence of light neutral CP-odd or CP-even scalar particles below the mass of the 
observed Higgs boson\cite{Cacciapaglia:2016tlr}. 
A CP-even scalar can for example be identified as the radion mode present in warped extra-dimension models with bulk gauge fields. 
A CP-odd scalar is typically a pseudo Nambu Goldstone boson from an approximate global symmetry, just like those appearing in 
composite Higgs models. The couplings to gauge fields are induced by the many fermion resonances populating the TeV scale 
(see e.g \cite{Belyaev:2016ftv} or also \cite{Fichet:2016xvs}).

As part of the 2017 Les Houches workshop on TeV scale 
physics~\cite{Brooijmans:2018xbu}, a simplified model was used to examine whether measurements at the LHC can give 
information about such possible particles. Here we present updated results of that study.

The study uses an effective theory approach to describe a scalar with mass $M_\phi$ interacting with gauge bosons. 
The effective theory has $SU(2)\times U(1)_Y$ symmetry.
When the scalar is light, with a mass below the electroweak scale, we assume that it has large tree-level 
$SU(2)\times U(1)_Y$ couplings, so that the loop-induced electroweak-breaking contributions are subleading. 
Under these conditions the interactions of a CP-even and CP-odd scalars with gauge bosons are described by 
dimension-5 effective Lagrangians. 
Mixing with the SM Higgs is assumed to be small to ensure that the SM Higgs has SM-like couplings compatible with 
measurements. A common scale $\Lambda$ for all couplings was assumed. More details are given in~\cite{Brooijmans:2018xbu}.

The measurements of interest are those involving isolated photons, or pairs of photons, in the final state. 
These have been measured inclusively~\cite{Aad:2012tba,Aad:2013zba,Aad:2016xcr}, and in 
association with jets~\cite{Chatrchyan:2013mwa,Aad:2013gaa,ATLAS:2012ar}, $W$ or $Z$ bosons\cite{Aad:2013izg,Aad:2016sau} 
(i.e. leptons and/or missing energy). 
At low $M_\phi$ and low-ish $\Lambda$, one of the most sensitive measurements is the $\gamma+E_T^{\rm miss}$ measurement 
from \cite{Aad:2013izg}. 
The Higgs fiducial diphoton measurements~\cite{Aad:2014lwa} are also of interest. 
These were studied and in principle 
have some sensitivity -- events generated by the models considered do contribute to the fiducial region. However, since 
the value of $M_\phi$ considered here lie below the SM Higgs mass, the events which will enter the fiducial phase space
of the Higgs measurement will arise from combinatorial backgrounds of pairs of photons, and thus will not exhibit a 
peak at the Higgs mass. Because of this, they are likely to removed as part of the background fitting and subtraction 
process in that analysis. We therefore do not include the Higgs cross sections when calculating the sensitivity.

A study carried out around the same time as the Les Houches study~\cite{Mariotti:2017vtv} using lower energy data as well as LHC data, 
but not the LHC boson+$\gamma$ data, obtains similar results, although due to differences in the definition of the couplings a precise
comparison has not been carried out.

\begin{figure}[h]
\includegraphics[width=0.9\textwidth]{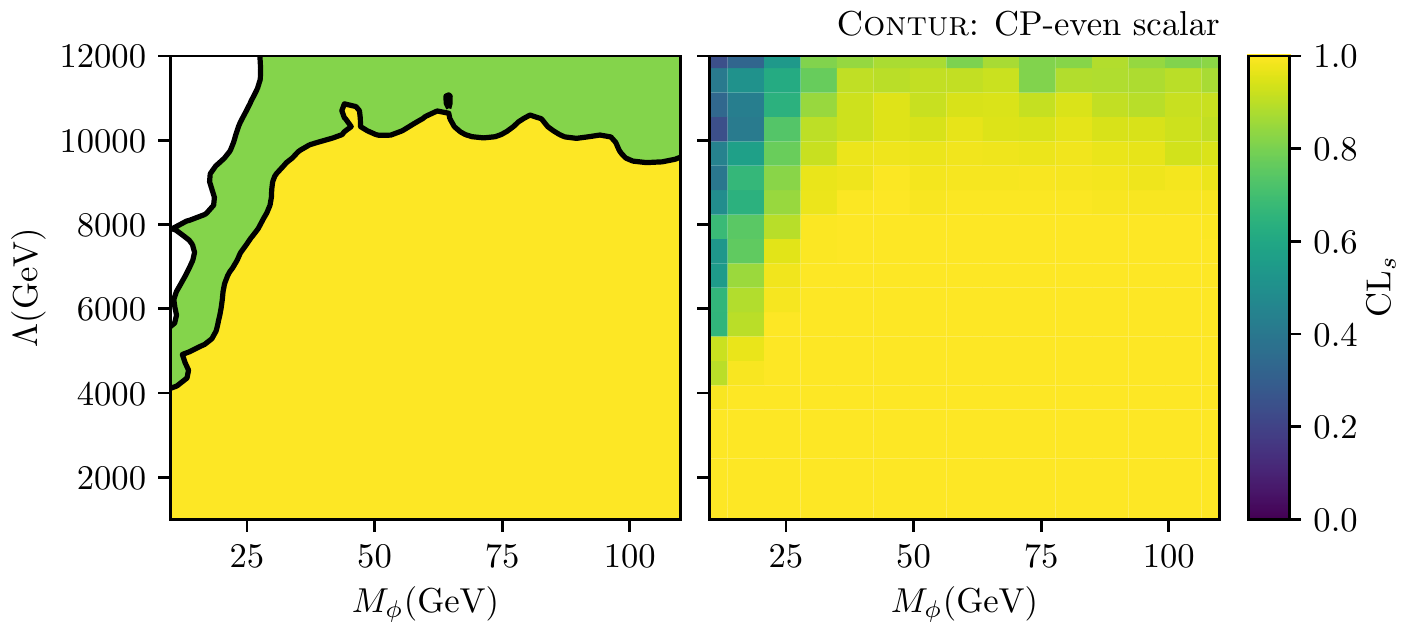}
\caption{\label{fig:cpe} 
Sensitivity of LHC measurements to a light CP-even scalar particle $\phi$ decaying to photons.}
\end{figure}

\begin{figure}[h]
\includegraphics[width=0.9\textwidth]{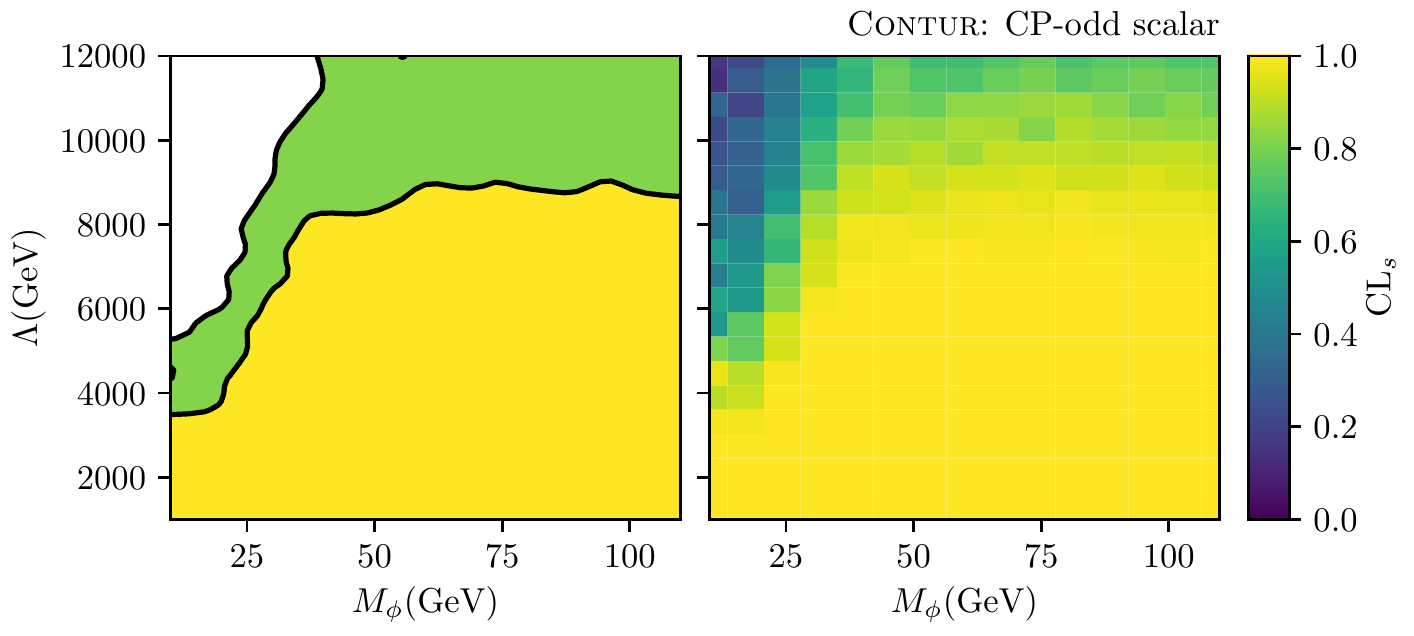}
\caption{\label{fig:cpo} 
Sensitivity of LHC measurements to a light CP-odd scalar particle $\phi$ decaying to photons.}
\end{figure}

Figure~\ref{fig:cpe} shows the LHC sensitivity in the $\Lambda$ vs $M_\phi$ plane for the CP-even scalar.
Figure~\ref{fig:cpo} the equivalent sensitivity for the CP-even scalar.

Dependent on $M_\phi$, the $\Lambda$ values up to 4.5 to 10.5~TeV are excluded, under the assumptions of our procedure, 
for the CP-even scalar. For the CP-odd scalar, the maximum reach is around 8.5~TeV in $\Lambda$. 

\section{Summary and Future Plans}

The \contur approach exposes significant sensitivity to SM extensions in unfolded particle-level measurements. Where
dedicated searches for these models have already been performed, see \cite{Whalen,Pandolfi}, \contur has similar sensitivity 
if the same data set is used; 
however, in several cases it lags behind either because the measurements have not yet been made, or because they have not been 
made available in HEPDATA and Rivet. For BSM scenarios which have not been considered by dedicated searches, \contur provides
an efficient way to identify models and regions of parameter space which are disfavoured by existing data, 
and those which remain of interest. The ability to consider the exact, inclusive phenomenology of a new model means that
\contur smoothly transitions between different signals as the parameters of the model change the dominant processes.

At present, \contur does not make use of the full correlation information which is available for several of the measurements considered.
The intention is to do so, which should increase the sensitivity is several cases. Also, in these studies we have taken the
data -- which are consistent with the SM -- to be exactly equal to the SM, looking for whether the BSM contributions 
would have made a visible difference compared to the experimental uncertainties. 
In this mode, \contur can only ever exclude BSM scenarios, and where 
the SM theory uncertainty is large, it may in fact overestimate the exclusion. 
It is however possible to make use of precision final-state SM calculations, which means that in future \contur could potentially
identify BSM scenarios which describe the data {\it better} than the SM.

More results and updates are available at https://contur.hepforge.org .

\section*{Acknowledgments}
This work has received funding from the European Union's Horizon 2020 research and innovation programme as part of the 
Marie Sklodowska-Curie Innovative Training Network MCnetITN3 (grant agreement no. 722104).
I'd like to thank my \contur co-authors, especially David Yallup, as well 
as Peter Richardson who provided impressive support with the Herwig UFO interface. Finally, I would like to thank
the conference organisers for the invitation and for hosting a really stimulating meeting, and especially
Deepak Kar for his hospitality at University of the Witwatersrand.

\section*{References}

\bibliographystyle{iopart-num}
\bibliography{Contur-Kruger}

\end{document}